\documentclass[graybox]{svmult}

\usepackage[utf8]{inputenc}
\usepackage{mathptmx}       
\usepackage{helvet}         
\usepackage{courier}        
\usepackage{type1cm}        
\usepackage{makeidx}         
\usepackage{graphicx}        
\usepackage{multicol}        
\usepackage[bottom]{footmisc}
\usepackage{amsfonts}
\usepackage{amsmath}
\usepackage{makecell}

\makeindex             

\begin{document}

\title*{Unsupervised Network Embedding for Graph Visualization, Clustering and Classification}
\author{Leonardo Guti\'errez G\'omez \and Jean-Charles Delvenne}
\institute{Leonardo Gutierrez Gomez \at Institute for Information and Communication Technologies, Electronics and Applied Mathematics (ICTEAM) \at Université catholique de Louvain, Avenue Georges Lemaitre, 4, 1348 Louvain-la-Neuve, Belgium  \email{leonardo.gutierrez@uclouvain.be}
\and Jean-Charles Delvenne \at Institute for Information and Communication Technologies, Electronics and Applied Mathematics (ICTEAM) and Center for Operations Research and Econometrics (CORE) \at Université catholique de Louvain, Avenue Georges Lemaitre, 4, 1348 Louvain-la-Neuve, Belgium  \email{jean-charles.delvenne@uclouvain.be}}
%
%
\maketitle

\abstract*{A main challenge in mining network-based data is finding effective ways to represent or encode graph structures so that it can be efficiently exploited by machine learning algorithms. Several methods have focused in network representation at node/edge or substructure level. However, many real life challenges such as time-varying, multilayer, chemical compounds and brain networks involve analysis of a family of graphs instead of  single one opening additional challenges in graph comparison and representation.
Traditional approaches for learning representations relies on hand-crafting specialized heuristics to extract meaningful information about the graphs, e.g statistical properties, structural features, etc. as well as engineered graph distances to quantify dissimilarity between networks. \newline\indent
In this work we provide an unsupervised approach to learn embedding representation for a collection of graphs so that it can be used in numerous graph mining tasks. By using an unsupervised neural network approach on input graphs, we aim to capture the underlying distribution of the data in order to discriminate between different class of networks.
Our method is assessed empirically on synthetic and real life datasets and evaluated in three different tasks: graph clustering, visualization and classification. Results reveal that our method outperforms well known graph distances and graph-kernels in clustering and classification tasks, being highly efficient in runtime.}

\abstract{
A main challenge in mining network-based data is finding effective ways to represent or encode graph structures so that it can be efficiently exploited by machine learning algorithms. Several methods have focused in network representation at node/edge or substructure level. However, many real life challenges such as time-varying, multilayer, chemical compounds and brain networks involve analysis of a family of graphs instead of  single one opening additional challenges in graph comparison and representation.
Traditional approaches for learning representations relies on hand-crafting specialized heuristics to extract meaningful information about the graphs, e.g statistical properties, structural features, etc. as well as engineered graph distances to quantify dissimilarity between networks. \newline\indent
In this work we provide an unsupervised approach to learn embedding representation for a collection of graphs so that it can be used in numerous graph mining tasks. By using an unsupervised neural network approach on input graphs, we aim to capture the underlying distribution of the data in order to discriminate between different class of networks.
Our method is assessed empirically on synthetic and real life datasets and evaluated in three different tasks: graph clustering, visualization and classification. Results reveal that our method outperforms well known graph distances and graph-kernels in clustering and classification tasks, being highly efficient in runtime.}

\section*{Introduction}
Numerous complex systems in social, medical, biological and engineering sciences can be studied under the framework of networks. Network models are often analyzed at the node/edge or substructure level, studying the interaction among entities, identifying groups of nodes behaving similarly or finding global and local connectivity patterns among a given network. Furthermore, many real life challenges might involve collections of networks representing instances of the system under study, e.g functional brain networks (connectomes) \cite{10.1371/journal.pbio.0060159}, chemical compound graphs \cite{Srinivasan:1997:PTE:1624162.1624163}, multilayer networks \cite{DBLP:journals/corr/abs-1212-2153}, and so on. Other applications involve dynamic interactions between components, introducing an additional complexity in the time evolution of the system. For example, in a social mobile phone network, people are considered as nodes and the phone calls as edges. The dynamics of calls between users will systematically add and remove edges between them, describing a sequence of static graphs characterizing a dynamic evolution of the system. 
 
With the increasing availability of manually labeled network data, many of these problems  have recently raised the attention of the machine learning community.  Machine learning applications seek to make predictions or discovering patterns in graph structured data. For example, in chemoinformatics \cite{doi:10.1021/jm00106a046}, one might need to predict the toxicity or anti-cancer activity of proteins and molecules represented as graphs. In time-varying social networks, one might be interested in detecting unusual events \cite{Peel:2015:DCP:2888116.2888122}, e.g points in time in which the network connectivity differs abruptly with respect to the evolution of the underlying process. Prediction of subjects having a neural disorder such as Alzheimer or Schizophrenia, based on their connectomes is crucial in neuroscience \cite{doi:10.1002/hbm.22633}. 

The cornerstone of this approach is the feature representation of the input data, e.g finding effective ways to encode graph structures in such a way that it can be used in traditional machine learning models. For example, in order to predict whether a molecule is toxic or not, one might build a feature vector representation of a molecule incorporating information about its atoms, as well as global and local properties of the graph structure itself \cite{Barnett2016, doi:10.1093/comnet/cny034}. By doing so we can train a traditional machine learning model such as support vector machines, random forest, neural network, etc. so it will discriminate unseen toxic and non-toxic chemical compounds. 

There exist many manners to extract features and comparing networks. For instance, graph distances \cite{2018arXiv180107351D, Livi:2013:GMP:2737203.2737238} such as the Jaccard and Hamming distances compute differences between graphs by counting the number of edit operations to transform a graph into another one, focusing mainly in their local connectivity patterns. Other distances are spectral in nature based on the comparison between the eigenvalues of the reference matrices representing the networks. Another popular class of distance measures are the graph kernels \cite{Shervashidze:2011:WGK:1953048.2078187, Yanardag:2015:DGK:2783258.2783417}. A kernel can often be seen as the scalar product between implicit high-dimensional feature representations of a network  \cite{973}. The so-called kernel trick allows to compare networks without ever computing explicitly the coordinates of data points in the high-dimensional feature space, sometimes with a substantial gain in computational time over classical graph-distance approaches.
 
\begin{figure}[!t]
\includegraphics[scale=0.53]{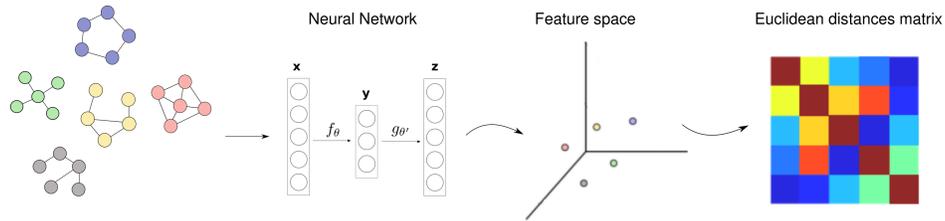} 
\caption{Overview of the proposed method. Given a family of graphs, we train an unsupervised neural network in order to uncover dissimilar relationships between graphs. The graphs are embedded into a feature space and mapped to a Euclidean distance matrix reflecting the structural similarity between input examples.}\label{general_pipeline}
\end{figure}

However, real life networks are complex structures involving heterogeneous connectivity patterns across domains, constraining the expressiveness of the aforementioned methods in multiple tasks. Therefore, the most relevant hand-crafted features tend to be task dependent and often require specialized domain expertise in order to incorporate the right properties to perform accurately on the target task. 

Unlike previous approaches, in this work we propose a method to learn network embeddings from a collection of networks. It should not be confused with node embedding approaches which aim to map nodes from a graph into vectors on a feature space (see \cite{DBLP:journals/corr/abs-1709-07604, DBLP:journals/corr/GoyalF17} for a survey of those methods). Therefore, in this paper we refer to graph or network embedding the outcome of mapping each network of a family as a vector in a Euclidean space (see Figure \ref{general_pipeline}). The unsupervised nature of the method allows to learn the most relevant features from the data in order to produce lower dimensional representation of input graphs. This reduces the curse of dimensionality of high dimensional graphs uncovering discriminative relationships in the underlying dataset. As a consequence, networks with similar structural properties will have neighboring embeddings in the feature space, and dissimilar graph will be more distant. Our approach thus differs from the various definitions of graph distances or similarities mentioned previously in that we learn automatically a feature representation of graphs assessing their similarity on a Euclidean space, instead of using a hand-crafted metric in the graph space. In addition, because many graph created in real life applications rarely have exchangeable nodes, we focus on problems defined on networks that account for node identities, e.g time-varying networks, brain networks, multilayer networks, etc.

We evaluate our method empirically in three network mining tasks: graph clustering (grouping similar graphs together), graph classification (predicting the class to which unseen networks belong to) and visualization (plotting many networks in $\mathbb{R}^2$).
We perform diverse experiments on synthetic and real life datasets such as time-varying networks (primary school network),  multilayer networks (European airport network) and brain networks datasets.

This paper is structured as follows. First, we introduce some popular methods of the literature used to compare networks, as well as the development of the proposed approach. Then, we present some applications in graph visualization, clustering and classification performed on synthetic and real life datasets. Subsequently, a computational analysis of our method is presented, finalizing with a discussion and perspectives for future work.

\section*{Methods}
\subsection*{\textbf{Graph distances}}\label{sec_graph_dist}

Distinguishing among a class of networks requires a notion of distance or similarity between pairs of graphs \cite{2018arXiv180107351D}. These measures capture different aspects of the local and global structure of graphs having an impact in the outcome of different applications. We present some of the most representative graph distances of the literature. 

The \textit{Hamming} and \textit{Jaccard} distances are special instances from the broader class of graph-edit distances. They measure the number edge insertion and deletion operations necessary to transform one graph to another one. Denoting $N$ the number of nodes of the undirected graphs $G_1=(V,E_1)$ and $G_2=(V,E_2)$ with adjacency matrices $A_1$ and $A_2$ respectively, the \textit{Hamming} distance between them is defined as:

\begin{equation}
d_H(G_1,G_2) = \dfrac{1}{N(N-1)}\sum_{i,j}^N|A_1 - A_2|_{i,j} 
\end{equation}
which defines a scaled version of the $L_{1,1}$ norm between matrices bounded between $0$ and $1$.
Similarly, the \textit{Jaccard} distance is defined as:

\begin{equation}
d_J(G_1,G_2) = \dfrac{|E_1 \cup E_2| - |E_1 \cap E_2|}{|E_1 \cup E_2|} 
\end{equation}
where $E_1$ and $E_2$ are the set of edges for the graphs $G_1$ and $G_2$ respectively.\\

\textit{DeltaCon} \cite{Koutra:2016:DEC:2888412.2824443} is a popular graph similarity measure in connectomics. As the edit distances it also exploits node correspondence across graphs. The intuition behind the method is to compute first pairwise node similarities of input graphs through a variant of a personalized PageRank algorithm \cite{Koutra:2016:DEC:2888412.2824443}. The pairwise node affinity matrices $(S_1, S_2)$ are compared using the Matusita Distance defined by:

\begin{equation}
d_{DC}(S_1,S_2) = \sqrt{\sum_{i,j=1}^n (\sqrt{S_1(i,j)} - \sqrt{S_2(i,j)})^2}
\end{equation} 

On the other hand, the \textit{spectral distances} for graphs have proven to be very useful in many applications \cite{10.1038/s41598-018-37534-2, Wilson:2008:SGS:1379924.1380381}. However, the spectral nature of the method makes it invariant to node permutations. Roughly speaking, these methods compare the spectrum of any matrix representing the input graph, generally the graph Laplacian. The combinatorial Laplacian matrix (CL) of an undirected graph $G$ is defined by $L = D - A$, where $D$ is the diagonal matrix whose $i$-th element equal to the degree of node $i$, and $A$ its adjacency matrix. The normalized Laplacian matrix (NL) is defined by $L^{'} = D^{-1/2}L D^{1/2} = I - D^{-1/2}AD^{1/2}$, with $I$ the corresponding identity matrix. We denote the eigenvalues of any of the Laplacian matrices as $0=\lambda_1 \leq \lambda_2 \leq  \ldots \leq \lambda_N$. 

For any $L$ and $L^{'}$ we consider the following spectral distance. The spectral distance between two undirected graphs $G_1$ and $G_2$ is defined as \cite{10.1038/s41598-018-37534-2}:

\begin{equation}
d(G_1, G_2) = \sqrt{\sum_{i=1}^{n_{\lambda}} [\lambda_{N+1-i}(G_1) - \lambda_{N+1-i}(G_2)]^2}
\end{equation}

where $n_{\lambda}$ is the number of eigenvalues considered for the computation of the distance (typically $n_{\lambda}=N$). 

\subsection*{\textbf{Embedding distances}}
Unlike the previous distances, our approach performs network comparisons directly on a feature space through a learned non-linear mapping applied to input graphs (see Figure \ref{general_pipeline}). The building blocks of our method are explained in the following subsections.

\subsubsection*{\textbf{Autoencoder}} 
Unsupervised learning approaches aim to uncover hidden patterns or learning representations from unlabeled data. The autoencoder (AE) \cite{Vincent:2008:ECR:1390156.1390294} is one of the most popular unsupervised neural network approaches. It has been widely used as a performant mechanism to pre-train neural networks and general purpose feature learning \cite{DBLP:journals/ijon/ChoiCB18}. It allows to compress the representation of input data, disentangling the main factors of variability, removing redundancies and reducing the  dimension of the input.

Given a set of data examples $D = \{\textbf{x}^{(1)},\textbf{x}^{(2)},\ldots , \textbf{x}^{(m)} \}$, the purpose of the the traditional auto-encoder is to learn a non-linear mapping which encodes an input example $\textbf{x} \in \mathbb{R}^n$ in a smaller dimensional latent vector $\textbf{y} \in \mathbb{R}^d$ with $n \gg d$. The encoding mapping has the form of $f_\theta(\textbf{x}) = s(W\textbf{x} + b)=\textbf{y}$, generally through a non-linear function $s$ such as sigmoid or $tanh$ applied entrywise on the vector $W\textbf{x}+b$. A reverse mapping of $f$ is used to reconstruct the input from the feature space: $g_{\theta'}(\textbf{y}) = s(W' \textbf{y} + b')=\textbf{z}$. The parameters  $\theta = \{W,b\}$ and $\theta' = \{W',b'\}$ are optimized by minimizing the average reconstruction error over the training set:

\begin{equation}\label{loss_ae}
\theta^*, \theta'^* = \underset{\theta, \theta'}{arg \min} \hspace*{0.2cm} \dfrac{1}{m}\sum_{i=1}^m \| \textbf{x}^{(i)} - \textbf{z}^{(i)} \|_2^2
\end{equation}

Note that when $s$ is the trivial identity, the solution is equivalent to the classical PCA (principal component analysis) with the number of hidden units as the principal components. One can therefore see autoencoders as a nonlinear extension of PCA.

\begin{figure}[!t]
\begin{center}
\includegraphics[width=2.5in]{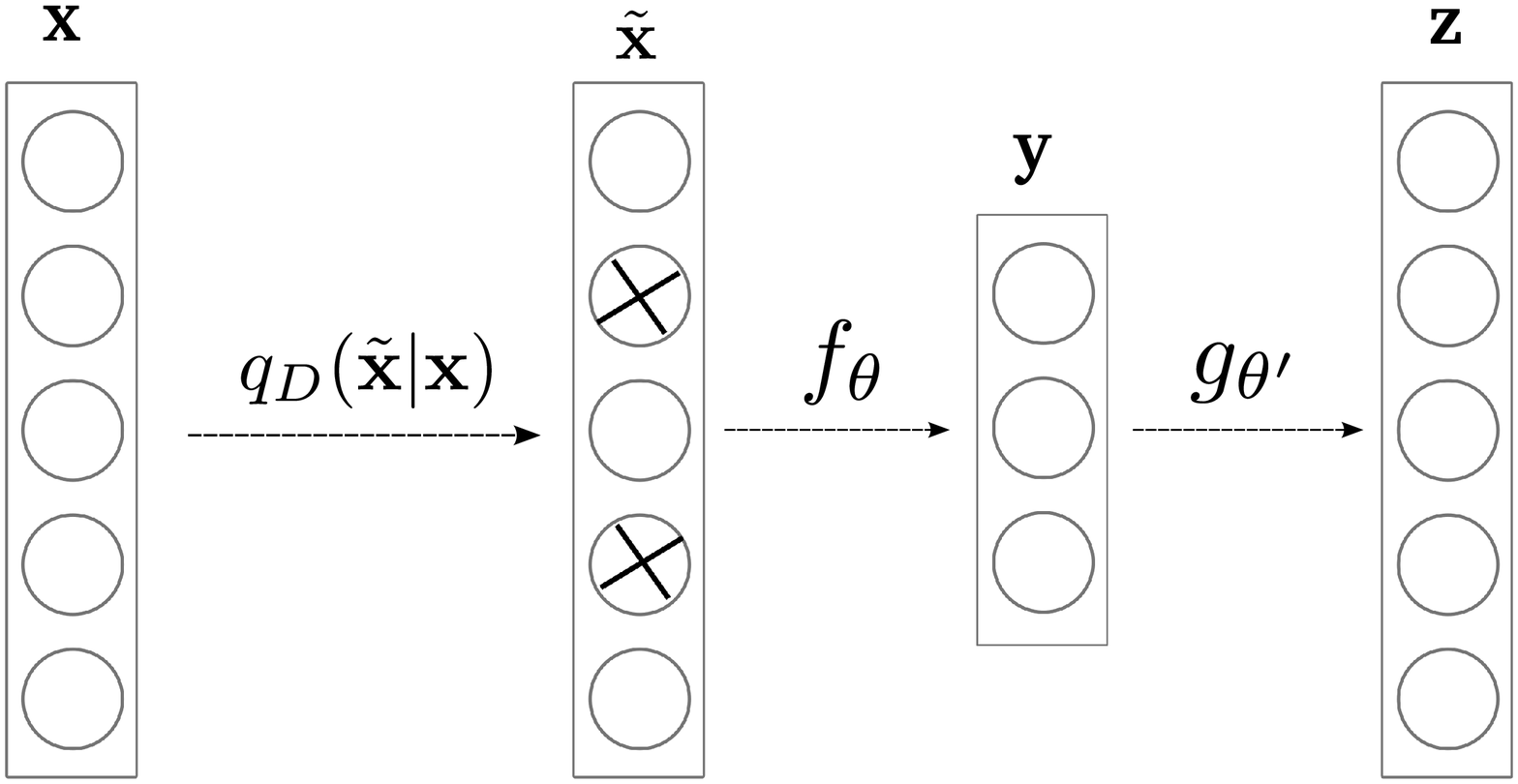} 
\caption{Denoising Autoencoder. A corrupted instance \textbf{$\tilde{x}$} of a graph \textbf{x} is fitted into the Autoencoder's input. The Autoencoder is trained to recover a cleaned version of the input by compressing it through a non-linear mapping $f_{\theta}$ and mapping it back (through $g_{\theta'}$) to a reconstructed version of the original input graph \textbf{x}.}\label{DAE}
\end{center}
\end{figure}

\subsubsection*{\textbf{Denoising Autoencoder (DAE)}}\label{sec_DAE}
Minimizing the previous reconstruction criterion alone is unable in general to guarantee the extraction of meaningful features as it can potentially memorize the training data. We want the Autoencoder to be sensitive enough to recreate the original observation but insensitive enough to the training data such that the model learns a generalizable encoding and decoding mapping. 

To avoid this limitation, the objective (Eq \ref{loss_ae}) is redefined in such a way that the autoencoder will be able to clean partially corrupted input or simply denoising it. This modification leads a simple variant of the basic autoencoder described above. A denoising autoencoder (DAE) \cite{Vincent:2008:ECR:1390156.1390294} is trained to reconstruct a clean or repaired version from a corrupted input. This is done by transforming the original input $\textbf{x}$ in $\tilde{\textbf{x}}$ through a stochastic mapping $\tilde{\textbf{x}} \sim q_D(\tilde{\textbf{x}}|\textbf{x})$. By doing so the AE is forced to learn meaningful features, robust under corruption of the input. 

The corrupted version $\tilde{\textbf{x}}$ is mapped with the original autoencoder to a hidden representation $\textbf{y}=f_{\theta}(\tilde{\textbf{x}})$ from which we reconstruct a clean $\textbf{z} = g_{\theta'}(\textbf{y})$. An important observation is that $\textbf{z}$ is now a deterministic function of $\tilde{\textbf{x}}$ rather than $\textbf{x}$. See Figure \ref{DAE} for an schematic representation of the model. Thus, we optimize the same objective than Eq \ref{loss_ae} but replacing $\textbf{x}$ by $\tilde{\textbf{x}}$. Optimization is done with the standard mini-batch gradient descent and back propagation algorithms \cite{Lecun1998}.

\subsubsection*{\textbf{Network embedding distances} }
The adjacency matrix $A$ of a graph is a simple network representation but alone can be insufficient as an input for the DAE. It only captures first order relationships between neighboring nodes. We extend this by computing higher powers of the adjacency matrix in order to capture multiple paths relationships. Thus, we consider $A^r$ for some $ r\geq 1$ as a more adequate input for the Denoising Autoencoder. 

Note that as the class of problems we tackle are defined on a collection of networks having a node correspondence across graphs, our method remains invariant to the node ordering when the same node permutation is assigned to the graphs.

The vectorization of matrices is required to feed the graphs into the DAE input. Let $A^r$ the $r$ power of the  $n \times n$ adjacency matrix $A$ of a graph. The vectorization of $A^r$ is a $n^2 \times 1$ column vector $\textbf{x}=vec(A^r)$ obtained by staking the columns of $A^r$. Notice that when the graph is undirected, the input matrix can be described with a $\frac{n(n + 1)}{2} \times 1$ column vector $\textbf{x}$. We apply a stochastic noise on the input by removing or adding an small fraction of edges at random, then we infer the parameters of the DAE using the noisy inputs $\tilde{\textbf{x}}$ as was presented in the previous section.

The optimal solution $\theta^* =\{W^*,b^*\}$ parametrizes an encoder mapping $f_{\theta^*}$ of the DAE. It embeds the input $\textbf{x} = vec(A^r)$ into a smaller dimensional vector $f_{\theta^*}(\textbf{x}) \in \mathbb{R}^d$.
A main advantage of transforming graphs into feature vectors is that it allows us to compare easily networks computing only Euclidean distances between their embeddings. Hence, the \textit{network embedding distance} between two graphs $G_1$ and $G_2$ with power matrices $A_1^r$ and $A_2^r$ is defined as:

\begin{equation}\label{emb_dist}
d(G_1, G_2) = \| f_{\theta^*}(vec(A_1^r)) - f_{\theta^*}(vec(A_2^r))\|_2.
\end{equation} 
In the following sections we present some experimental results of our method in various synthetic and real life  applications.

\section*{Experiments and Results}

The experiments have three purposes. First, they assess the performance of our method in discriminating different types of networks which are generated from different models, edge densities and heterogeneous community structure.
Next, they show the use of graph embeddings in networks coming from diverse real life applications such as time-varying networks, connectomes and multilayer networks. Finally, they highlight the runtime performance of feature computation and compare it against other techniques. It is worth to mention that all our experiments were performed with $A^3$ as input for the DAE.
  
We evaluate our approach on three different but related tasks: graph visualization, graph clustering and classification. A detailed report of the parameters used in our experiments can be found in the appendix.

\begin{figure}[t]
\includegraphics[width=4.9in]{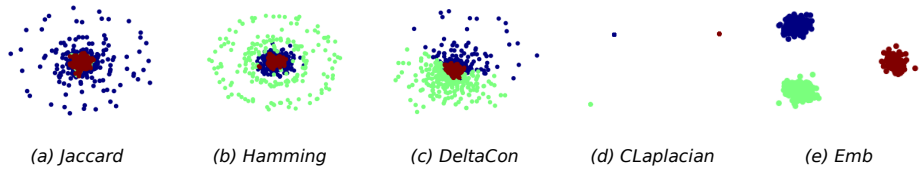}
\caption{Visualization of permuted Erd\H{o}s-Rényi (ER) dataset. Each point corresponds to a network. Color of a point indicates the category of the network according with its average degree $\langle d \rangle$. For blue $\langle d \rangle=4$, green $\langle d \rangle$ = 6 and red color $\langle d \rangle=8$.}\label{many_viz}
\end{figure}

\subsection*{\textbf{Graph visualization}}
A useful application of network embedding is graph visualization. It mainly consists in representing graph as 2D points, e.g an entire graph as one point, maximizing a certain notion of similarity. Considerable research has been done in visualizing nodes of graphs based on the premise that nodes sharing common structures e.g. neighboring nodes, structural equivalent nodes, assortative nodes, etc. should be mapped to close points in the embedding space \cite{DBLP:journals/corr/abs-1709-07604, DBLP:journals/corr/GoyalF17}. 

In contrast, we propose to visualize multiple graphs at once on a two-dimensional space in the following way. From a given family of graphs, their embeddings are learned and used to compute the embedding distance matrix (Eq. \ref{emb_dist}). In order to enable a visualization, a methodology is needed to bring the embedding distances into a low-dimensional visualization. We choose the Multi-scale SNE tool \cite{Lee2015} as standard method. This is a non-linear dimensionality reduction approach for data visualization which aims to reproduce in a low-dimensional space the local and global neighborhood similarities observed on any similarity matrix. In this way, we expect that networks with similar properties as learned by the Denoising Autoencoder are neighboring points in the two dimensional visualization, while the gap between dissimilar groups of graphs is maximized.

\subsubsection*{\textbf{Visualizing synthetic networks} }\label{sec_viz}
To assess the relevance of the visualization, we generate synthetic random networks with a range of parameters, and assign a color to each point in the visualization that reflects the value of the parameter used to generate the network. In this way we expect that a good visualization will preserve the same colored points as neighbors points in $\mathbb{R}^2$ maximizing the gap between groups. We generate two synthetic datasets which are described in the following.

\subsubsection*{\textbf{Datasets}} In the first synthetic dataset, we create three Erd\H{o}s-Rényi networks (\textit{ER}) with different parameters (Figure \ref{many_viz}). Then we generate $200$ copies from each graph reordering the nodes with a different permutation of the original graph. In the second dataset, $1000$ power law networks were generated using the Lancichinetti–Fortunato–Radicchi (\textit{LFR}) benchmark \cite{Fortunato:2009:CDA:1698822.1698858}. This algorithm creates networks with heterogeneous structures and communities sizes. The mixing parameter $\mu \in [0,1]$ controls the strength of the community arrangements, achieving well defined communities with small $\mu$, meaningless community structure when $\mu$ is close to one and $\mu=0.5$ as the border beyond which communities are no longer defined in the strong sense \cite{Radicchi2658}. Thus, we generate two groups of networks: one with mixing parameter $\mu=0.1$ and other with $\mu=0.5$. Other parameters are common for both groups: number of nodes $N=81$, average node degree equal $11$, community sizes varying between $6$ and $22$ nodes, exponent for the degree sequence $2$ and exponent for the community size distribution $1$.
Therefore, the two groups of networks differ only in the strength of their communities structure and not in the degree distribution, being a more challenging problem than the previous dataset. \\

\subsubsection*{\textbf{Discussion} } Figure \ref{many_viz} shows the visualization of the \textit{ER} dataset after applying our method and the aforementioned graph distances. As can be expected, results with \textit{Jaccard} and \textit{Hamming} distances are not satisfactory because points from different groups overlap. Even though \textit{DeltaCon} tries to separate the data, the boundary between groups was not clearly determined. Because spectral distances are permutation invariant measures, they collapse all permuted graphs to the same point showing a hard separation between classes. On the other hand, our embedding (\textit{Emb}) shows three well defined cloud of points grouping together isomorphic graphs. Our method exploits node correspondence across graphs when it is known, but even if we lose track of node order we can retrieve networks that are essentially identical.

\begin{figure}[h]
\begin{center}
\includegraphics[width=2.0in]{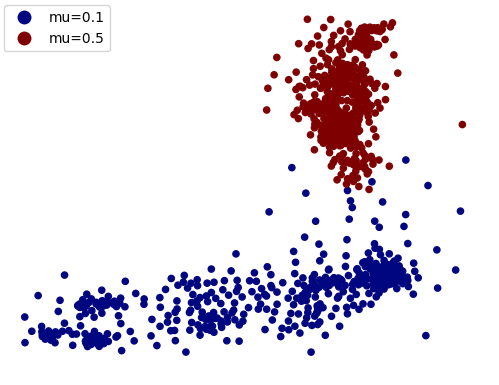}\hspace*{1cm} 
\includegraphics[width=2.1in]{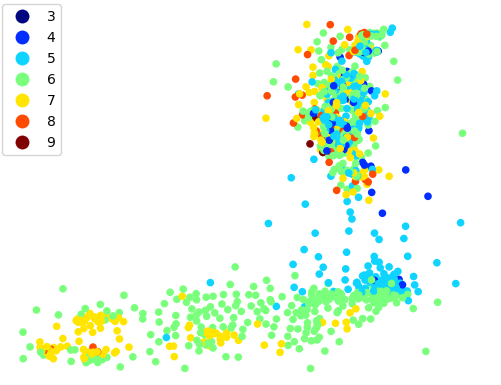} 
\end{center}

\caption{Visualization of networks generated with the LFR benchmark. \emph{(Left)} networks with different community strength: $\mu=0.1$ and $\mu=0.5$. \emph{(Right)} Same left-hand side networks. Colors encode the number of planted communities within each network.}\label{dataset_viz}
\end{figure}

Figure \ref{dataset_viz} shows the visualization of the \textit{LFR} dataset. The left-hand side plot shows two clouds of points encoding networks with different mesoscopic structure. As can be seen the blue cluster tends to spread more than the red one, which is more compact. This illustrates the structural variability of networks having heterogeneous number/size communities (blue cluster) against a group of networks with weakly modularity  (red cluster). 

In the right-hand side plot of Figure \ref{dataset_viz}, we keep the same networks from the left-hand side plot, but we color them according with the number of ground truth communities on each network.
Inspecting the bottom cluster we observe that even if there is not a clear grouping of points, the data is distributed in a quasi-continuum manner, having networks with similar number of communities as neighboring points in the plane. On the other hand, the group of networks on the top are indistinguishable, which is expected because their weak community strength. This visualization allows us to understand the notion of similarity captured by the Autoencoder on the underlying dataset.

Once more we emphasize that although the embedding is in principle dependent on the order of the nodes, in this specific case different orderings lead to closely similar visualizations. This is expected as in this case, albeit all the networks are supported on the same number of nodes, there is no natural one-to-one correspondence between the nodes of two networks, and all nodes are treated symmetrically in the generation process. 

\subsubsection*{\textbf{Visualizing real life networks: temporal networks}}
The primary school network \cite{10.1371/journal.pone.0023176} is a dataset containing temporal face-to-face interactions between 232 children and 10 teachers in a primary school in Lyon, France. The data was collected over two days (Thursday, October 1 and Friday, October 2, 2009) spanning from 8:45 am to 5:20 pm the first day, and 8:30 am to 5:05 pm the second day. 

\begin{figure}[h]
\begin{center}
\includegraphics[width=2.8in]{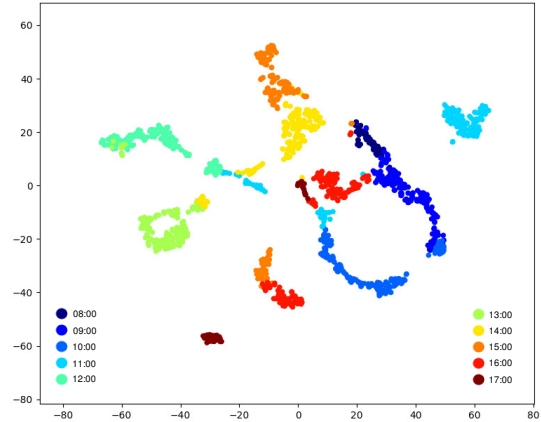} 
\caption{Visualization for primary school embeddings. Each point represents a network snapshot during the day of the 01-October-2009. Color of points encode a time frame of the day spanning from 8:45 until 17:20}\label{school}
\end{center}
\end{figure}

The dynamic evolution of the network can be modeled as a time-varying network defined on a fixed number of nodes, and dynamic edges representing the physical interaction between children and teachers. It can be represented as a sequence of static graph snapshots over a time window $\tau$ which aggregates all events or edge activations occurred between the interval $[(t-1)\tau ,t\tau]$. For this experiment, we chose a time resolution $\tau = 20s$ yielding 1230 snapshots for Thursday, 01-October. Its visualization is shown in Figure \ref{school}.

The clusters in Figure \ref{school} can be seen as groups of networks behaving similarly and correlated with external events, e.g consecutive clusters are separated because an external event. For instance, lunch time is characterized by clusters defined between 12:00 and 14:00. The class time is represented by a long cluster of dark and light blue points in the morning and yellow, orange groups in the afternoon. The end of the school day is highlighted with a brown group of points. Mixed group colors indicates smooth temporal transitions, e.g. end of lunch time and beginning of classes (green-yellow-light blue), also the end of the afternoon break to classes (orange-red).

Note that unlike synthetic examples from the previous section, nodes have an individual identity, and different network snapshots take place on the same set of nodes.

\subsection*{\textbf{Graph clustering}}

\subsubsection*{\textbf{Clustering synthetic graphs}}
Another important application is clustering of networks. Clustering aims to group together  ``similar'' graphs and putting dissimilar ones in different groups. We proceed alike the previous section, but we do not perform dimensionality reduction to $\mathbb{R}^2$. Instead, clustering is performed directly in the embedding space with the standard spectral clustering algorithm \cite{NIPS2001_2092}. This technique makes use of the spectrum of a similarity matrix of the data to performing clustering in fewer eigenvectors.

We create four different synthetic datasets composed by 600 networks of 81 nodes each. We run our method on each dataset and compute a $600 \times 600$ network embedding matrix using Eq \ref{emb_dist}. In order to compare against other techniques, the graph distance matrices for the methods introduced in the first part of the manuscript are computed. All matrices are normalized having a maximum value of one for dissimilar pairs of graphs and zero for the most similar ones. Therefore, spectral clustering is performed on the similarity matrices induced by the previous graph distance measures.  

The clustering performance is evaluated through the normalized mutual information (NMI) \cite{10.1007/978-3-642-14366-3_23} metric in the form:

\begin{equation}
NMI(C_t, C_p) = \dfrac{2I(C_t;C_p)}{H(C_t)+H(C_p)}
\end{equation}
where $H$ is the entropy of a class distribution and $I$ the mutual information between the ground truth class distribution $C_t$, and the predicted cluster assignment $C_p$. It runs from zero when the algorithm fails to a value of one when the clustering is perfectly recovered. Details about the datasets and ground truth class generation are presented in the following.

\begin{table}
\caption{Summary of sythetic datasets} 
\label{datasets}
\begin{tabular}{p{2cm}p{3.0cm}p{4.9cm}p{2cm}}
\hline\noalign{\smallskip}
\textbf{DATASET} &\textbf{Type of network} &\textbf{Properties} & \textbf{True clusters} \\ 
\noalign{\smallskip} \hline\noalign{\smallskip}
ER & Erd\H{o}s-R\'enyi & Different average degrees & 4 \\ [2pt]
Mixed & ER - Power law & Different models, same average degree & 2 \\ [2pt]
LFR & Power law  &  Strong vs weak communities strength & 2 \\ [2pt]
Dynamic & Erd\H{o}s-R\'enyi &  Perturbation mechanism: rewiring, adding and removing \% edges & 3 \\ [2pt]
\noalign{\smallskip}\hline\noalign{\smallskip}
\end{tabular} 
\end{table}

\subsubsection*{\textbf{Datasets}} An overview of the generated datasets is shown in Table \ref{datasets}. We generate Erd\H{o}s-Rényi (\textit{ER}) networks with four distinct parameters producing random networks with different average degrees. The so called \textit{Mixed} dataset is a collection of power-law networks generated by the Barabási-Albert (\textit{BA}) model and  Erd\H{o}s-Rényi networks, all with the same average degree. 

In an attempt to simulate a dynamic network evolution, we simulate a time varying network (\textit{Dynamic}) following \cite{2018arXiv180107351D}, applying a perturbation mechanism from a starting ER network. At each time step, a fraction of edges of the previous graph are rewired uniformly at random. At the same time, we apply a depletion/thickening process in which edges are deleted with probability $0.015$ and formerly absent edges are added with probability $0.015$. We introduce two perturbation points by augmenting the probabilities of adding and deleting edges to $0.2$ from time $t=200$ and also to $0.6$ from time $t=400$, defining three ground truth clusters of similar behaving networks. Finally, the \textit{LFR} dataset introduced previously for networks visualization is also considered for clustering. For all datasets we generate balanced ground truth classes.  
 
In order to evaluate the sensitivity of the clustering to the node ordering, we perform clustering with different enumeration of nodes by applying a fixed node permutation across the networks. We reported the mean and standard deviation of the $NMI$ after running the experiment ten times.

\subsubsection*{\textbf{Discussion}}
Regarding the clustering results in Table \ref{clustering}, we observe that our graph embeddings (\textit{Emb}) provides better clustering than traditional graph distances. The method is capable to differentiate networks with different edge densities (\textit{ER}). Meanwhile, it is able also to discriminate networks with different degree distribution even if they have a similar average degrees (\textit{Mixed}).  Discriminating power law networks from strong to weak community structure (\textit{LFR}) is also well achieved. The  time-evolving network (Dynamic) is a harder setting in which our method perform the best comparatively to graph distances. In this case the graph embeddings are able to capture the variations introduced by anomalous points in the underlying evolution of the network. This can be explained because the DAE was not designed for a target kind of graphs. Instead, it learns the underlying distribution of the data, identifying the main factor of variability adapting its parameters for discriminating heterogeneous networks. The quality of the embeddings remains almost the same after permuting the nodes, which is confirmed by the low variance in the NMI. Hence, in practice we fixed a node numbering for the learning procedure.

\begin{table}[h]
\caption{Clustering results for synthetic datasets (NMI)} \label{clustering}
\begin{tabular}{l@{\quad}c@{\quad}c@{\quad}c@{\quad}c@{\quad}c@{\quad}l}
\hline\noalign{\smallskip}
  & \textbf{Hamming} & \textbf{Jaccard} & \textbf{DeltaCon} & \textbf{CLP} & \textbf{CLP normed}& \textbf{Emb} \\ 
\hline\noalign{\smallskip}
ER	&	0.024	&	0.070	&	0.294	&	\textbf{0.933}	&	\textbf{0.914}	&	\textbf{0.918 $\pm$ 0.004} 	\\ [2pt]
Mixed &	\textbf{1.0	}    &	\textbf{1.0}	    &	\textbf{1.0	}    &	 \textbf{1.0}    &	0.374  &\textbf{ 1.0 $\pm$ 0}   \\ [2pt]
LFR	&	0.219	&	0.603	&	0.265	&	0.035	&	\textbf{0.983}  	&	\textbf{0.986 $\pm$ 0.014}	\\ [2pt]
Dynamic	& 0.389	&	0.255	&	0.198	&	0.216	&	0.172	&	\textbf{0.652 $\pm$ 0.085}	\\ [2pt]
\hline\noalign{\smallskip}
\end{tabular} 
\end{table}

\subsubsection*{\textbf{Clustering real life networks: multilayer networks}}
\begin{figure}[!h]
\includegraphics[scale=0.45]{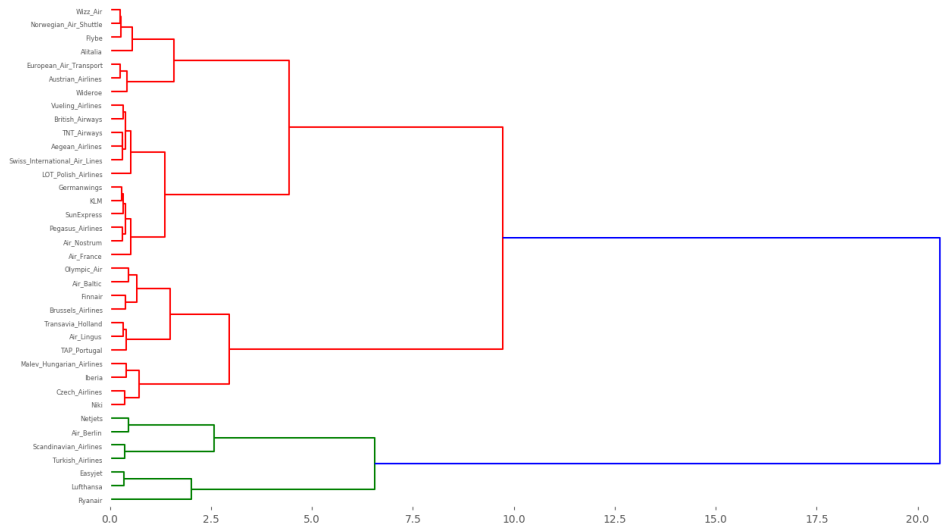} 
\caption{Dendrogram of airlines for European airports}\label{dendrogram}
\end{figure}

The European Air Transportation Network (ATN) \cite{DBLP:journals/corr/abs-1212-2153} is a multilayer network with 37 layers each representing a different European airline. Each layer has the same number of nodes which represent 450 European airports. We learn graph embeddings for all layers and we cluster them applying a standard hierarchical clustering algorithm on the network embedding distance matrix.
The hierarchical clustering provides partition of layers according with their similarity on the embedding space, see Figure \ref{dendrogram}.

Our findings confirm those introduced in \cite{DBLP:journals/corr/abs-1212-2153}. We can identify two main clusters representing major and low-cost aerial companies, as well as some regional airlines grouped together. Indeed, these airlines have developed according with different structural/commercial constraints. Low-cost companies tends to avoid being centralized and cover more than one country simultaneously. Major airlines have a hub and spoke network, connecting outlying airports to few central ones, providing a maximum coverage from their home country.

\subsection*{\textbf{Graph Classification}}
We evaluate graph classification in the context of supervised classification. 
It requires previously annotated reference samples (graphs) in order to train a classifier and subsequently classify unknown data.

\subsubsection*{\textbf{Brain connectomes classification}}
In this experiment we apply our method on a brain networks (connectomes) dataset  built from magnetic resonance imaging (MRI)~\cite{B.Chiem2018}. 
Structural and diffusion MRI data of 91 healthy men and 113 healthy women is preprocessed in order to create undirected networks. All graphs have the same 84 nodes representing neural Regions of Interests (ROIs). Weighted edges correspond to the number of neural fibers linking two ROIs. The ROI keeps the same correspondence among graphs. The task is to classify connectomes according to gender,  male or female.

\subsubsection*{\textbf{Experimental setup}}
We assess the performance of our method against some well known algorithms for graph classification, mainly graph kernels and feature-based approaches. We choose the Shorthest Path (SP) and the Weisfeiler-Lehman (WL) subtree kernels \cite{Shervashidze:2011:WGK:1953048.2078187}. We also compare against the feature-based (FB) method \cite{Barnett2016} and Multi-hop assortativities features (MaF) for network classification \cite{doi:10.1093/comnet/cny034}. Such methods provide a pairwise similarity matrix between networks in the form of a Gram matrix which is used to train a popular support vector machine classifier (SVM) \cite{Smola:2004:TSV:1011935.1011939}. 
Note that the graph distances considered in this work do not define a proper positive semi-definite matrix. Therefore, following \cite{Wu05ananalysis} we shift the spectrum of their similarity matrices providing a proper kernel coherent with the SVM setting.

We follow the experimental setup of \cite{Shervashidze:2011:WGK:1953048.2078187, Yanardag:2015:DGK:2783258.2783417}. The dataset is randomly split in training and testing sets. The best model is cross-validated over 10 folds. Parameters of SVM are optimized only
on the training set. Thus, we compute the generalization accuracy on the unseen test set. In order to exclude the random effect of the data splitting, we repeated the whole experiment 10 times. Finally, we report the average prediction accuracies and its standard deviation.

For each graph kernel we report the result for the parameter that gives the best classification accuracy. For the feature-based approach \cite{Barnett2016}, feature vectors were built with the same network features they reported in their paper: number of nodes, number of edges, average degree, degree assortativity, number of triangles and global clustering coefficient. Results are shown in tables \ref{brains1} and \ref{brains2}.

\begin{table}[h]
\caption{Mean and standard deviation of classification accuracies on brain connectomes dataset.}\label{brains1}
\begin{tabular}{c@{\quad}c@{\quad}c@{\quad}c@{\quad}c@{\quad}l}
\hline\noalign{\smallskip}
\textbf{WL} & \textbf{SP} & \textbf{FB}& \textbf{CLaplacian}& \textbf{NLaplacian}& \textbf{Emb} \\ 
\hline\noalign{\smallskip}
$61.20 \pm 2.16$	& $65.45 \pm 1.78$	& $65.95 \pm 2.54$    &	$74.19 \pm 11.16$ & $71.07 \pm 10.95$ & $\textbf{87.20} \pm \textbf{7.60}$\\[2pt]
\hline\noalign{\smallskip}
\end{tabular} 
\end{table}
\begin{table}[h]
\centering
\caption{Mean and standard deviation of classification accuracies on brain connectomes dataset.}\label{brains2}
\begin{tabular}{c@{\quad}c@{\quad}c@{\quad}c@{\quad}c@{\quad}l}
\hline\noalign{\smallskip}
\textbf{Hamming} & \textbf{Jaccard} & \textbf{DeltaCon} & \textbf{MaF} & \textbf{Emb} \\ 
\hline\noalign{\smallskip}
$84.37 \pm 9.26$	& $84.34 \pm 10.11$	& $\textbf{87.80} \pm\textbf{ 6.54}$	& $84.26 \pm 5.81$    & $\textbf{87.20} \pm \textbf{7.60}$\\[2pt]
\hline\noalign{\smallskip}
\end{tabular} 
\end{table}

{\scriptsize{*Bold values correspond to the most performing techniques}}

\subsubsection*{\textbf{Discussion}} As can be seen in Table \ref{brains1}, WL, SP and FB perform significantly worse than spectral distances and graph embedding. This is expected as they do not take the identity of the nodes into account. Here, all brains share the same anatomical regions, which make the order of the nodes relevant. 
In Table \ref{brains2} can be seen that among the approaches exploiting node correspondence, our method (Emb) outperforms all others while remaining competitive with DeltaCon.

\subsection*{\textbf{Computational cost}}
Our graph embedding approach involve globally two steps: learning graph embedding through the DAE followed by a pairwise Euclidean distance matrix computation (Eq \ref{emb_dist}). In order to make fair comparisons, for each method of Table \ref{datasets} we measure the runtime for computing the distance matrix between pairs of graphs over synthetic datasets. The running times reported for our approach include learning graph embeddings and pairwise Euclidean distances computation. Results are shown below in Figure \ref{runtime}.

\begin{figure}[!ht]
\begin{center}
\includegraphics[width=3.6in]{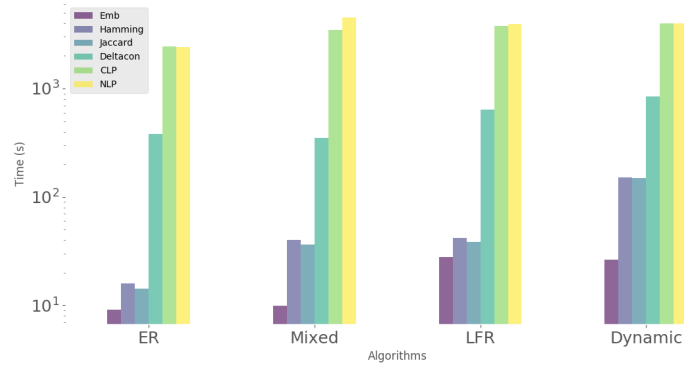} 
\caption{Computational time for feature computation. Time is log scaled.}\label{runtime}
\end{center}
\end{figure} 
As can be seen our method (Emb) outperform all graph distances across all studied datasets. The competitor approaches compute their similarity score comparing examples directly in the graph domain. However, we compare graphs in the embedding space. It is well known that spectral distances (CLP, NLP) are heavy in computation due to the eigenvalues calculation. Hamming and Jaccard distances rely in computing common node/edges patterns being slower in dense networks. Even if Deltacon is a scalable graph similarity measure, it is outperformed by the edit distances, but is more efficient than spectral distances. Meanwhile, our graph embedding method remains the fastest. Indeed, mini-batch gradient descent on relatively small datasets convergences faster. For efficiency reasons, the Euclidean distance between two feature embeddings $x,y$ was computed as $d(x,y) = \sqrt{\langle x, x\rangle - 2\langle x,y \rangle + \langle y, y \rangle}$. This formulation has the advantage of being very efficient for sparse graphs given that some terms can be pre-computed for an entire pairwise computation.

All computations were done on a standard computer Intel(R) Core(TM) i7-4790 CPU, 3.60GHzI with 16G of RAM.

\section*{Discussion and concluding remarks}
In the presented work we propose a method to learn graph embeddings for a collection of networks, e.g mapping graphs to $\mathbb{R}^p$ vectors. Our method allows to compare graphs computing only Euclidean distances between their embeddings in a feature space. We evaluate our method in three different applications in graph clustering, visualization and classification. Across heterogeneous synthetic and real life datasets, we compare our approach against well known graph distances and graph-kernel methods of the literature. 

It turns out that our approach extract the most appropriated features for distinguish different kind of graphs. Indeed, clustering groups of similar networks provides good quality partitions among synthetic datasets (Table \ref{clustering}), discriminating better heterogeneous structures among networks. Despite there is not a clear agreement about the use of combinatorial or normalized Laplacian in graph mining applications, spectral distances are highly competitive in graph clustering and visualization but are incapable to exploit node correspondence. Nevertheless, our learned graph embeddings turns out to be computationally cheaper than all considered methods (Table \ref{runtime}), being an attractive yet efficient method for comparing networks.

The results in graph classification reveal that our approach has superior performance than graph-kernels and graph spectral distances (Table \ref{brains1}). Indeed, exploiting the node identities across graphs increases the accuracy of the method. Thus, this result suggest a promising research direction in the connectomics domain.

Note that in this work we were not focusing in the task of for instance differentiating random networks with different average degrees, which can be trivially solved without any machine learning tool. Instead, we aimed to show an automatic way to leave the machine figure out the most relevant hidden patterns from the data, which is more general than designing tailored methods for particular applications.
   
The current study was limited by the assumption that all networks must have the same set nodes. Even if in many real applications this hypothesis holds, a large amount of complex systems have heterogeneous size graphs, e.g. chemical compounds, social networks, etc. This study has only investigated the class of graphs without node/edge attributes, such as age, gender in social networks. Addressing these issues introduce additional challenges and new opportunities for further research.

Despite this limitation, our work has the potential of being extended in two directions. Because the DAE captures the underlying probability distribution of the data \cite{Vincent:2008:ECR:1390156.1390294}, the decoding function could be used to generate artificial data, e.g generating brain networks, for mining purposes. Another possibility is to explore deeper neural network architectures such as the stacked autoencoders \cite{Vincent:2010:SDA:1756006.1953039} and its variants in order to learn hierarchical feature representation of the data for graph classification and clustering applications.

\begin{acknowledgement}
This work was supported by Concerted Research Action (ARC) supported by the Federation Wallonia-Brussels Contract ARC 14/19-060; Fonds de la Recherche Scientifique; and Flasgship European Research Area Network (FLAG-ERA) Joint Transnational Call "FuturICT2.0". We also thank Leto Peel and Michel Fanuel for helpful discussion and suggestions.
\end{acknowledgement}
%

%

\bibliographystyle{plain}
\bibliography{graph_class}

\end{document}